\documentclass[12pt]{article}
\usepackage{epsf,epsfig,amsmath}
\begin{document}

\title{On Questions of Non-Locality in our EPR Model}

\author{Karl Hess$^1$ and Walter Philipp$^2$}

\date{$^1$ Beckman Institute, Department of Electrical Engineering
and Department of Physics,University of Illinois, Urbana, Il 61801
\\ $^{2}$ Beckman Institute, Department of Statistics and Department of
Mathematics, University of Illinois,Urbana, Il 61801 \\ }
\maketitle

\begin{abstract}

The Einstein-locality of our EPR model has been questioned. We
show that our model obeys Einstein locality and that the questions
are without basis.

\end{abstract}

Questions have been raised whether our model for EPR experiments
\cite{hpp1}, \cite{hpp2}, \cite{hpp3} contains violations of
Einstein locality. We respond here to these questions by Myrvold
\cite{myr} and by Gill, Weihs, Zeilinger and Zukovsky (GWZZ)
\cite{gwzz}, \cite{gwzz1}. Their arguments are along similar lines
and we therefore attempt to answer their questions together.

We point to the fact that time-related parameters may be
correlated over space-like distances without violating Einstein
locality. This fact complicates the necessity to define Einstein
locality in mathematical terms. A question of particular
importance is whether joint distributions of random variables that
characterize EPR models may depend on the settings of both
stations when Einstein locality is postulated. Our answer is:
definitely yes, they may, provided that certain conditions are
fulfilled which can easily be met when time-related parameters are
involved. We assume that the reader is familiar with our previous
work and notation \cite{hpp2} and just repeat briefly the most
essential aspects of our EPR model.

The basic assumption of our model is that time and setting
dependencies of hidden instrument related variables are of the
essence. It is claimed by GWZZ in references \cite{gwzz} and
\cite{gwzz1} that time is irrelevant. Thus, GWZZ need to show that
time and time dependencies are of no concern in EPR experiments as
they are currently implemented. To show this, they cite a thought
experiment ``repeating the measurement procedure....not as a
sequence of successive repetitions at the same locations, but in a
million laboratories all over the galaxy". They state that they
compare ``potential outcomes under (sic) different settings at the
same time". Thus their thought experiment considers measurements
all over the galaxy all at the same time. It is clear that such a
thought experiment cannot have any relation to an actual
experiment. Simultaneous measurements all over the galaxy are
neither well defined from the viewpoint of the theory of
relativity nor quantum mechanics. In addition, it has been
enunciated by Peres \cite{peresn} that a one shot experiment that
leads to the Bell contradiction must be considered ``sheer
nonsense". Even if a large number of certain experiments could be
performed at a given instant of time, such experiments would not
be equivalent to sequential EPR experiments as they are currently
performed \cite{gwzz} if time and setting dependent instrument
parameters (as we have described them) are involved. The question
remains then whether time dependencies can exist in EPR
experiments as they are actually performed. To our knowledge this
question has not been answered in the literature. Therefore we can
assume as a working hypothesis that such time dependencies
actually exist. Under such a hypothesis, theories such as that of
GWZZ that do not consider time dependencies can not describe the
experiments. GWZZ claim that their thought experiment represents
local realism and that the Bell inequalities follow from this
local realism. This claim represents, within the framework of our
hypothesis of time dependent EPR experiments, just a logical
circle. We will show this in more detail below.

Before doing so, however, we need to clarify a misinterpretation
of GWZZ. GWZZ quote our paper \cite{hpp2}:

``Label the corresponding functions $A$ and $B$ as $A_{(m)}$ and
$B_{(m)}$ and consider the index $(m)$ a function of the source
parameter $\lambda = ({{\lambda}^1}, {{\lambda}^2})$ and the time
operators $O_{{\bf a},t}^1$, $O_{{\bf b},t}^2$. Then the functions
$A_{(m)}$ and $B_{(m)}$ can be considered as functions of $\bf a$,
$\lambda$, ${{\Lambda}_{{\bf a},t}^1}$, and $\bf b$, $\lambda$,
${{\Lambda}_{{\bf b},t}^2}$, respectively."

Now GWZZ note that the index $m$ depends on the settings of both
sides and that therefore the functions $A, B$ which are indexed by
$m$ will depend on both settings. We have stated explicitly before
\cite{hp} that the random variables need to be considered
mathematically a function of all the variables including those of
both stations in order to define the integrals that appear in the
theory. However, they actually depend $\bf {only}$ on the
variables of the respective side. We have emphasized over and over
that $A, B$ depend on the local variables only. We also have never
used or implied any dependency of the functions $A, B$ on the
settings of the other side as can easily be seen from our papers
\cite{hp}, \cite{hpp2}. Nevertheless, we admit that the paragraph
above should more properly have been formulated as follows:

``Label the corresponding functions $A$ and $B$ as $A_{(m)}$ and
$B_{(m)}$ and consider the index $(m)$ a function of the source
parameter $\lambda = ({{\lambda}^1}, {{\lambda}^2})$ and the time
operators $O_{{\bf a},t}^1$, $O_{{\bf b},t}^2$. Then the functions
$A_{(m)}$ and $B_{(m)}$ can be considered as functions of $m$
where, in addition, $A$ depends $\bf {only}$ on $\bf a$,
$\lambda$, ${{\Lambda}_{{\bf a},t}^1}$, and $B$ depends $\bf
{only}$ on $\bf b$, $\lambda$, ${{\Lambda}_{{\bf b},t}^2}$,
respectively."

We apologize, if our original formulation was not clear enough.
The space provided in \cite{hpp2} was simply not enough to explain
all the details of our model. We are currently preparing a more
extensive description of our model in easier mathematical terms.
In this more extensive version, the index $m$ simply represents a
time interval in which just one correlated pair is measured and
the settings are chosen after the choice of $m$. The parameter
$\lambda$ is also separated from $m$ \cite{mathmod}.

We turn now to the central argument of Myrvold \cite{myr} which
does not relate to the functions $A, B$ but to the joint density
of our time and setting dependent instrument related parameters.
Myrvold claims that our model is nonlocal because the joint
density of our parameters ${{\Lambda}_{{\bf a},t}^1}$ at station
$S_1$ and ${{\Lambda}_{{\bf b},t}^2}$ at station $S_2$ depends on
both settings. Here Myrvold misses the central point. Because we
work with the hypothesis that there exist time dependencies and
time like correlations and because our parameters depend on the
respective setting and a global time $t$, the joint density must,
in general, depend on the settings. For example, it is easy to
show that two computers which are independent except for having
the same clock-time can create arbitrary parameters
${{\Lambda}_{{\bf a},t}^1}$ at station $S_1$ and ${{\Lambda}_{{\bf
b},t}^2}$ at station $S_2$ with probability densities that depend
on both settings just as our probability densities do for a given
$m$ ($m$ corresponding now to a given short time interval). With
setting dependent joint densities, as we have described them, the
Bell inequalities can not be proven \cite{foot0}. This shows the
logical circle of Myrvold and also GWZZ: they ignore time
correlations or pronounce time as irrelevant and therefore their
joint densities must be independent of the settings and the Bell
inequalities follow.

The only additional argument of GWZZ regards our index $i$ and is
just based on a lack of appreciation of our model, of basic
mathematical notation and of basic concepts in elementary
probability theory. In their Eq.[10] GWZZ claim that our function
$\kappa$ $\bf {does}$ depend on  $i$  and $j$. It $\bf {does}$
$\bf {not}$! The function $\kappa$ is simply the indicator
function of the union of unit squares, lined up along the main
diagonal of the big square $\Omega$: We have used an Einstein type
of convention to sum over doubly appearing indices. To make this
entirely clear, the indicator of the interval $[i-1,i)$ should
have been denoted by $1_{[i-1,i)}(u)$ as is standard notation in
mathematics and not by $1\{i-1 \leq u < i\}$, as we have done in
Eq.[26] of our paper \cite{hpp2}. Thus our application of the
Einstein summation convention in Eq.[26] should have been spelled
out and we apologize if this notational shortcut has caused a
misunderstanding \cite{foot1}. In addition, GWZZ fail to
appreciate the basic concept of a probability density with respect
to Lebesgue measure. Just as before $u$ and $v$ are the Cartesian
coordinates of a point $(u,v)$ in the Euclidean plane and not
local hidden parameters as GWZZ seem to surmise. Thus, for fixed
$\bf a$ and $\bf b$ the density  $\rho_{{\bf a}{\bf b}}$ only
depends on the Cartesian coordinates $u$  and $v$ and not on the
above index $i$. As a consequence of these two remarks it should
be evident that $i$ is not a random variable, as erroneously
claimed in GWZZ (their Eqs.[14] and [15]).

We conclude that the suspicions of GWZZ and Myrvold that our model
is nonlocal are without basis.

\end{document}